\documentclass[conference,letterpaper]{IEEEtran}

\usepackage[utf8]{inputenc} 
\usepackage[T1]{fontenc}
\usepackage{url}
\usepackage{ifthen}
\usepackage{cite}
\usepackage{graphicx}
\usepackage{xcolor}
\usepackage{array}
\usepackage[english]{babel}
\usepackage{amsthm}
\usepackage{amssymb}
\usepackage{multicol}
\usepackage{caption}
\usepackage{algorithm}
\usepackage{algpseudocode}
\usepackage{mathtools}
\usepackage{amsmath} 

\newtheorem{proposition}{Proposition}
\newtheorem{corollary}{Corollary}
\newtheorem{theorem}{Theorem}
\newtheorem{example}{Example}

\def\r{\mathbf{r}}
\def\s{\mathbf{s}}
\def\sgl{\sigma_{\mathrm{left}}}

\begin{document}
\setlength{\parfillskip}{0pt plus 1fil}
\title{Error-Resilient Weakly Constrained Coding via Row-by-Row Coding} 
\setlength{\abovedisplayskip}{3pt}
\setlength{\belowdisplayskip}{3pt}
\author{%
 \IEEEauthorblockN{Prachi Mishra \ and \ Navin Kashyap}
 \IEEEauthorblockA{Department of Electrical Communication Engineering\\
                   Indian Institute of Science,
                   Bengaluru, Karnataka, India\\
                    Email: \{prachimishra, nkashyap\}@iisc.ac.in}
                   
}



\maketitle


\begin{abstract}
 A weakly constrained code is a collection of finite-length strings over a finite alphabet in which certain substrings or patterns occur according to some prescribed frequencies. Buzaglo and Siegel (ITW 2017) gave a construction of weakly constrained codes based on row-by-row coding, that achieved the capacity of the weak constraint. In this paper, we propose a method to make this row-by-row coding scheme resilient to errors. 
\end{abstract}

\section{Introduction}
\label{sec:intro}
 In several applications where coding is required, it is necessary to prevent or restrict the frequency of occurrence of specific patterns as substrings of codewords. One such application area of much current interest is DNA-based data storage systems \cite{church2012},\cite{goldman2013}. DNA has great potential as a data storage medium because of its high density, physical compressibility, and stability over extreme environmental conditions. A DNA molecule is a polymer formed by a long chain of nucleotides. There are four types of nucleotides in such a chain, which means that DNA strands can be viewed a long sequences over a 4-letter alphabet. Repetitions of the same nucleotide, called the homopolymer run, significantly increase the chance of synthesis and sequencing errors \cite{bornholt2016dna}, \cite{ross2013characterizing}. Therefore, long homopolymer runs should be avoided in the design of DNA molecules for storage purposes. Other applications where it is necessary to restrict the frequencies of occurrence of certain patterns in codewords include the suppression of patterning effect due to inter-symbol interference \cite{shafarenko2010weakly},\cite{shafarenko2005skewed}, and multi-level cell flash memory \cite{liu2016shaping}.

A constrained code is a set of finite-length sequences that completely forbid certain unwanted patterns from appearing as substrings. A more relaxed notion is that of a weakly constrained code, which imposes restrictions on the frequencies of occurrence of certain patterns as substrings, without necessarily forbdding them completely. Weak constraints can be more appealing for applications such as DNA-based data storage, because strong constraints, which entirely prohibit the occurrence of undesirable substrings, often incur a significant rate penalty. There is a large body of literature on constrained codes (see e.g., \cite{MRS1998}, \cite{immink2004}, \cite{MRSlecturenotes}), but weakly constrained codes --- also called semi-constrained codes --- are relatively less explored \cite{marcus1992improved},\cite{immink1997weakly},\cite{elishco2016semiconstrained},\cite{buzaglo2017weakly}.

There is an extensive literature on making constrained codes error-resilient by combining them with error-correcting codes --- see \cite[Chapter~9]{MRSlecturenotes} for references. However, to the best of our knowledge, there seems to be no prior work on doing the same for weakly constrained codes. This paper attempts to address this problem. 

Specifically, we consider the weakly constrained coding scheme proposed by Buzaglo and Siegel \cite{buzaglo2017weakly}, which they have shown achieves the capacity of the associated weak constraint. In their coding scheme, messages are first encoded into a 2-dimensional array $W$ using row-by-row coding \cite{tal2009row}. A weakly constrained codeword $w$ is then formed by concatenating the columns of this array in some order. The order of concatenation of columns is not \emph{a priori} fixed, but depends on the messages being encoded. At the time of decoding, the decoder must first reconstruct the array $W$ from the codeword $w$. To do this, it must infer from $w$ itself the order in which the columns of $W$ were concatenated to form $w$. Errors affecting $w$ compromise the ability of the decoder to correctly determine the order of concatenation of columns. 

In this paper, we propose a modification to the coding scheme of \cite{buzaglo2017weakly}. Our scheme uses row-by-row coding to encode messages into arrays whose columns can always be concatenated in a \emph{fixed} order, while ensuring that the resulting codeword $w$ respects the weak constraint. This removes the need for the decoder to infer the order of concatenation from $w$, thus making the scheme more resilient to errors. Our modified scheme also introduces less redundancy than the original coding scheme of \cite{buzaglo2017weakly}. 

The remainder of this paper is structured as follows. We provide the necessary definitions and notation in Section \ref{sec:Preliminaries}. This section also contains a brief description of the weakly constrained coding scheme of \cite{buzaglo2017weakly}. We present our error-resilient weakly constrained coding scheme in Section~\ref{sec: Error-Resilient WeakCodes}. Some of the details of the coding scheme and its analysis are provided in appendices. We make some concluding remarks in Section~\ref{sec:Conclusion}. 
 
\section{Preliminaries}
\label{sec:Preliminaries}
This section contains only the definitions and notations required for our purposes. For details on how row-by-row coding schemes yield capacity-achieving coding schemes for weakly constrained systems, we refer the reader to the work of Buzaglo and Siegel \cite{buzaglo2017weakly}. 

A row-by-row coding scheme is associated with an $n$-integral Markov chain on a primitive subgraph of a de Bruijn graph. We introduce these notions below. For a positive integer $n$, we set $[n] := \{1,2,\ldots,n\}$. 

\subsection{Markov Chains}
Let $G = (V, E, L)$ be a labelled directed graph, where $V$ and $E$ denote the vertex and edge sets of $G$, respectively, and $L$ is the labelling function defined over the edges $e \in E$. For an edge $e\in E$, we let $\sigma(e)$ and $\tau(e)$ denote the initial and terminal vertex, respectively, of the edge.  A Markov chain $P$ on $G$ is a probability mass function over the edge set of $G$, $P: E \rightarrow [0,1]$, such that $\sum_{e \in E} P(e)=1 $. This induces a probability mass function, $\pi$, on the vertex set $V$: for all $u \in V$,
\begin{equation}
     \pi(u):=\sum_{e \in E : \sigma(e)=u}P(e).
     \label{eq:pi}
\end{equation}
A Markov chain is called \emph{stationary} if for all $ u \in V $, 
\begin{equation*}
     \pi(u)=\sum_{e \in E : \tau(e)=u}P(e).
\end{equation*}
Given a positive integer $n$, We call a stationary Markov chain \emph{$n$-integral} if, for all edges $e\in E$, the number $P(e)n$ is an integer.

\subsection{de Bruijn Graphs}
\begin{figure}[h]
    \centering
    \includegraphics[width=0.3\textwidth]{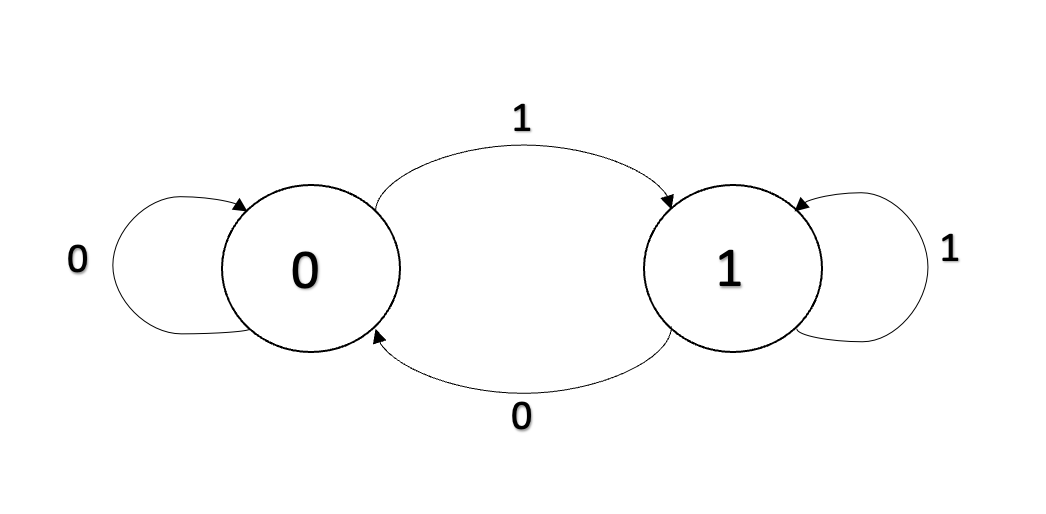}
    \caption{The de Bruijn graph $D_{1,2}$.}
     \label{fig:Graph}
\end{figure}

For integers $k \ge 1$ and $\mu \ge 2$, the $k^{\text{th}}$-order de Bruijn graph $D_{k,\mu}$ is a labelled directed graph over an alphabet $\Sigma$ of size $\mu$. Its vertex set $V_{k,\mu}$ is the set $\Sigma^{k}$. The edge set $E_{k,\mu}$ is defined such that there exists a directed edge from vertex $u=(u^1,u^2,\ldots,u^k)$ to $v=(v^1,v^2,\ldots,v^k)$ iff $u^{i+1}=v^{i}$ for all $i\in[k-1]$. An edge $e$ from $u$ to $v$ gets the label $L(e) := v^k$. Thus, the edge $e$ can be uniquely identified by a $(k+1)$-tuple $e=(e^{1},e^{2},\ldots,e^{k+1})$, where $e^{i}=u^{i}$ for all $i\in[k]$ and $e^{i+1}=v^{k}$. A first-order de Bruijn graph is shown in Figure~\ref{fig:Graph}.

A directed graph $G$ is said to be \emph{primitive} if it satisfies the following property for all sufficiently large integers $N$: for any ordered pair of vertices $u$ and $v$ in $G$, there is a path of length $N$ from $u$ to $v$. As an example, the primitive subgraphs of $D_{1,2}$ on the vertex set $V_{1,2} = \{0,1\}$ are precisely those obtained from $D_{1,2}$ by deleting at most one of the two self-loops. 

\subsection{Row-By-Row Constrained Coding} \label{sec:rowbyrow}

Let $P$ be an $n$-integral stationary Markov chain on a primitive subgraph, $G = (V,E,L)$, of $D_{k,\mu}$. We will assume that $P$ assigns positive probabilities to all the edges of $G$. Each edge $e$ of $G$ can then be viewed as having $P(e)n$ copies, which we call the \emph{multiplicity} of that edge. Listing out the vertices of $G$ in some arbitrary (but fixed) order $v_1, v_2, \ldots, v_{|V|}$, we define $n_{l} := \pi(v_{l})n$ for all $l \in [|V|]$, where $\pi(\cdot)$ is as defined in \eqref{eq:pi}. Note that $n_{l}$ is an integer for all $l$. We define a $k \times n$ matrix $U_\pi$ as follows: each of the first $n_1$ columns is equal to the vertex $v_1$, each of the next $n_2$ columns is equal to the vertex $v_2$, and so on, ending in the last $n_{|V|}$ columns all being equal to the vertex $v_{|V|}$.
Let $S(G)$ denote the set of all finite-length sequences over the alphabet $\Sigma$ that can be obtained by concatenating the labels of edges along (directed) paths in $G$. In other words, $S(G)$ consists of all possible words of the form $(L(e_1),L(e_2),\ldots,L(e_t))$, $t = 1,2,3,\ldots$, where the sequence of edges $e_1,e_2,\ldots,e_t$ forms a (directed) path in $G$.\\

For any $K \geq k+1 $, a \emph{$(G,P,n)$-array} is a $K \times n$ array $W$ having the following properties: 
\begin{itemize}
    \item[(A1)] each column of $W$ is a sequence in $S(G)$;
    \item[(A2)] in any $(k+1) \times n$ sub-array $W^{\#}$ formed by $k+1$ consecutive rows of $W$, each edge $e$ of $G$ occurs as a column of $W^\#$ exactly $P(e)n$ times.
\end{itemize}
Note that these properties imply that in any $k \times n$ sub-array $W^{\ddag}$ formed by $k$ consecutive rows of $W$, each vertex $v_l$ of $G$ occurs as a column of $W^{\ddag}$ exactly $n_l$ times.

A row-by-row constrained coding scheme is a means of encoding messages into a $(G,P,n)$-array.

\subsection{Weakly Constrained Coding} \label{sec:weakly}

 For a positive integer $N$, a \emph{$(G,P,N)$-weakly constrained coding scheme} is a method of encoding messages into (and subsequently decoding them from) length-$N$ words $w \in S(G)$ with the property that each edge $e$ in $G$ occurs in $w$ approximately $P(e)N$ number of times. 

Buzaglo and Siegel \cite{buzaglo2017weakly} proposed such a weakly constrained coding scheme using row-by-row coding. Their row-by-row coding scheme is an encoding of a sequence of $m$ messages into a $(G,P,n)$-array $W$ of size $(k+b\log n + m + N_G) \times n$, having some additional properties:
\begin{itemize}
\item[(B1)] the first $k$ rows of $W$ constitute the fixed matrix $U_\pi$ defined above;
\item[(B2)] the first $k+ b \log n$ rows of $W$ (where $b$ is some constant) constitute a fixed matrix $\hat{U}_\pi$ with all distinct columns;
\item[(B3)] the next $m$ rows below $\hat{U}_\pi$ encode the $m$ messages, one message per row;
\item[(B4)] the last $N_G$ rows (where $N_G$ is a constant that depends only on the graph $G$) are chosen so that the columns of $W$ can be stitched together in some order, to form a long word in $S(G)$. 
\end{itemize}

To satisfy Property (B3), the messages are first encoded into codewords of a certain constant-composition code $C_{\text{rbr}}$, and each codeword is then inserted in a prescribed manner into a row of the $(G,P,n)$-array $W$ --- see Appendix~\ref{sec:Crbr}.

Property (B4) requires further explanation. For two words $w_{1} = (w_1^1,w_1^2,\ldots,w_1^{\ell_1})$ and $w_2 = (w_2^1,w_2^2,\ldots,w_2^{\ell_2})$ over $\Sigma$, of lengths  $\ell_{1},\ell_{2} \geq k$, we say that $w_{1}$ is \emph{extendable} by $w_{2}$ if the last $k$ entries of $w_{1}$ are equal to the first $k$ entries of $w_{2}$. The result of extending $w_{1}$ by $w_{2}$ is word of length $\ell_{1}+\ell_{2}-k$:
\begin{equation}
    w_{1}||w_{2} := (w_{1}^1,w_1^2,\ldots,w_1^{\ell_1},w_2^{k+1},w_2^{k+2},\ldots,w_2^{\ell_2})
\label{eq:stitch}
\end{equation}
It can be readily verified that if two words $w_{1},w_{2} \in S(G)$ are such that $w_{1}$ is extendable by $w_{2}$, then $w_{1}||w_{2} \in S(G)$. Moreover, if an edge $e\in E$ occurs as a substring $t_1$ times in $w_{1}$ and $t_2$ times in $w_{2}$, then it occurs as a substring $t_1+t_2$ times in $w_{1}||w_{2}$. 

Now, let $W_1, W_2, \ldots, W_n$ be the $n$ columns of an $M \times n$ $(G,P,n)$-array $W$, listed in order from left to right. Suppose that there exists a permutation $\rho :[n] \rightarrow [n]$, such that for all $j\in[n-1] , W_{\rho (j)}$ is extendable by $W_{\rho (j+1)}$. Then,
$$
w=W_{\rho (1)}||W_{\rho (2)}||\cdots||W_{\rho (n)}.
$$
is a word of length $N = nM-(n-1)k$ belonging to $S(G)$, in which each edge $e$ of $G$ occurs exactly $P(e)n(M-k) = P(e) (N-k)$ times.

Lemma 2 in \cite{buzaglo2017weakly} says that if for all $e\in E$, we have $P(e)n\geq|V|$, then it is always possible to find some $N_{G}$ rows to satisfy Property (B4). More precisely, the addition of these $N_G$ rows yields a permutation $\rho:[n]\to[n]$ such that the $n$ columns of the resulting array $W$ can be stitched together as in \eqref{eq:stitch} to form a word of length $N=(b\log n + m + N_{G})n +k$ in which every pattern $e \in E$ occurs exactly $P(e)(N-k)$ times. We summarize this in the form of a proposition.

\begin{proposition}[Corollary~1 in \cite{buzaglo2017weakly}]
If for all $e\in E, P(e)n \geq |V|$ holds, then there exists a constant $b$ that depends only on $G$ and the Markov chain $P$, such that one can encode $m$ messages from $C_{\mathrm{rbr}}$ into a length $N=(b\log n +m+N_{G})n +k$ codeword $w\in S(G)$, in which every pattern $e\in E$ occurs exactly $P(e) (N-k)$ times.
\end{proposition}

This method of encoding $m$ messages into a length-$N$ codeword $w \in S(G)$ as above constitutes the encoder of a weakly constrained coding scheme. 

\begin{example}
\label{ex:graphG}
Let $G$ be the subgraph of $D_{1,2}$ shown in the figure below. 
\begin{figure}[hbt!]
    \centering
    \includegraphics[width=0.2\textwidth]{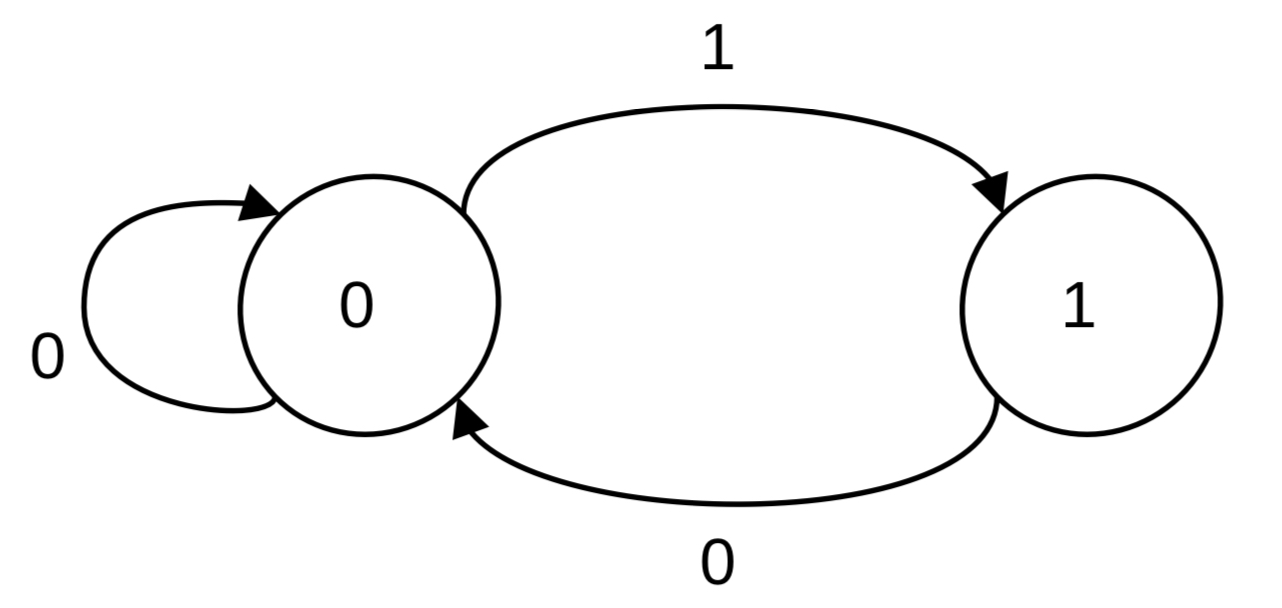}
\end{figure}
Consider the $n$-integral stationary Markov chain on $G$ given by $P(00)= 0.5, P(01) = P(10) = 0.25$, and $n=8$. For this graph and Markov chain, we give a $9 \times 8$ array $W$ that encodes $m=2$ messages  using the row-by-row coding scheme of Buzaglo and Siegel \cite{buzaglo2017weakly}:
\begin{center}
\begin{tabular}{ | m{0.5 cm} | m{0.5cm}| m{0.5cm} |m{0.5cm} |m{0.5cm} |m{0.5cm} |m{0.5cm} |m{0.5cm} | } 
  \hline
 0& 0& 0& 0& 0& 0& 1& 1 \\ 
  \hline
  0 & 0 & 0 & 0& 1& 1& 0& 0\\ 
  \hline
  0 & 0 & 0& 1& 0& 0& 1& 0 \\ 
  \hline
  0& 0& 1& 0& 0& 1& 0& 0 \\ 
  \hline
  0& 1& 0& 1& 0& 0& 0& 0 \\ 
  \hline
  1& 0& 0& 0& 0& 1& 0& 0 \\ 
  \hline
  0& 0& 0& 0& 0& 0& 1& 1 \\ 
  \hline
  0& 0& 1& 1& 0& 0& 0& 0 \\ 
  \hline
  1& 0& 0& 0& 0& 1& 0& 0 \\ 
  \hline
\end{tabular}
\end{center}

The first five rows of $W$ constitute $\hat{U}_{\pi}$ (of which the first row is $U_\pi$), followed by two rows representing messages, and the final $N_G=2$ rows needed to stitch together the columns. 
The columns can be stitched together in the order shown below to form a length-$65$ codeword $w \in S(G)$:
\begin{center}
$w=W_{1}||W_{7}||W_{2}||W_{3}||W_{4}||W_{5}||W_{6}||W_{8}$.\\
\end{center}
The codeword $w$ is of length $N=65$, in which the pattern $00$ occurs exactly $P(00)(N-k) = 32$ times, and the patterns $01$ and $10$ each appear exactly $P(01)(N-k) = P(10)(N-k)=16)=32$ times. 

\end{example}

Decoding requires recreating the $m$ rows of the array $W$ that constitute the \emph{payload}, i.e., the rows of $W$ that contain the messages. To do this, the decoder needs to know the order in which the columns of $W$ are stitched together to form the codeword $w$. It is able to infer this from the matrix $\hat{U}_\pi$, which is static and known to both encoder and decoder. The distinct columns of $\hat{U}_\pi$ act as unique identifiers for the columns of $W$. But this feature has the drawback that errors affecting the rows of $\hat{U}_\pi$ can potentially result in incorrect payload extraction by the decoder --- see Example~\ref{ex:Errors} in Appendix~\ref{ex: Error_ex}. On the other hand, any error in the $N_{G}$ rows added after the payload leads to an incorrect reconstruction of the $W$ matrix, but it does not affect the payload extraction. Therefore, if there is a way to avoid or correct the errors in the rows of $\hat{U}_{\pi}$, we can successfully extract the payload rows. Errors directly affecting the payload rows can be corrected using known error-correction techniques for constant-composition codes.

\section{Error-Resilient Weakly Constrained Coding}
\label{sec: Error-Resilient WeakCodes}
In this section, we propose a method to keep the order of concatenation of columns of $W$ constant, i.e., independent of the payload. In this way, we do not need the mechanism used in \cite{buzaglo2017weakly} of adding extra rows to create unique identifiers for the columns of $W$. This approach makes the construction of weakly constrained codes via row-by-row coding error-resilient. The number of redundant rows added by the method of \cite{buzaglo2017weakly} is $N_{G}+O(\log n)$, while our method adds only a fixed number of rows after the payload, independent of $n$. Our proposed method also removes the additional condition of $P(e)n \ge |V|$ required by the scheme in \cite{buzaglo2017weakly}. However, at this stage, our results are applicable only to primitive subgraphs $G$ of the first-order de Bruijn graph $D_{1,2}$.

\subsection{Our Main Results}

Let $G$ be a primitive subgraph of $D_{1,2}$, with vertex set $\{0,1\}$. Thus, $G$ is a graph obtained by removing at most one of the two self-loops in $D_{1,2}$. In particular, the edges $01$ and $10$ remain in $G$. Consider an $n$-integral stationary Markov chain on $P$ that assigns positive probability to all edges in $G$. Additionally, if $e$ is a self-loop of $D_{1,2}$ that is not in $G$, we set $P(e) = 0$. Stationarity of $P$ and primitivity of $G$ imply that $0 < P(01) = P(10) < 0.5$, and $\max\bigl(P(00),P(11)\bigr) > 0$.

\begin{figure}[t]
    \centering
    \includegraphics[width=0.5\textwidth]{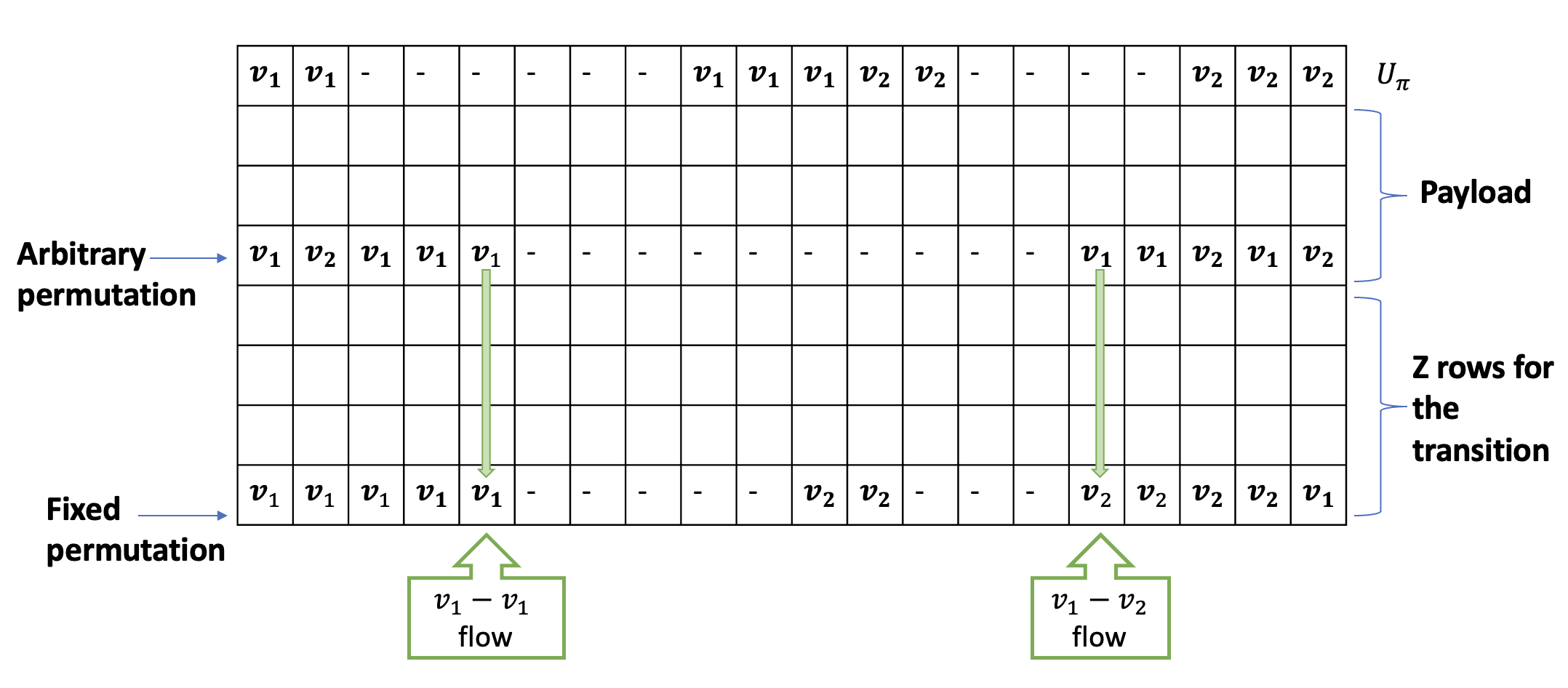} 
    \caption{A $(G,P,n)$-array of size $(1+m+Z) \times n$, with first row $U_\pi$, $m$ rows of payload, and last row $\sigma_{\mathrm{left}}(U_\pi)$.}
    \label{fig:Weakly Constrained Array}
\end{figure}

As in the scheme of \cite{buzaglo2017weakly}, we encode $m$ messages into a $(G,P,n)$-array $W$ using row-by-row coding. The first row of the array is the $1 \times n$ matrix $U_\pi$ which, as per its structure detailed in Section~\ref{sec:rowbyrow}, has vertex $v_{1} := 0$ appearing $n_{1}$ times as a block, followed by vertex $v_2 := 1$ appearing $n_2$ times. The next $m$ rows of $W$ constitute the payload of $m$ messages encoded using the code $C_{\text{rbr}}$, as described in Appendix~\ref{sec:Crbr}. At this stage, the $(m+1)$-th row of $W$ has the $n_1$ and $n_2$ copies of the vertices $v_1$ and $v_2$, respectively, arranged in an arbitrary order. Our goal is to transition from this arbitrary arrangement of vertices to a ``target'' row at the bottom of $W$ having a pre-determined arrangement of vertices, by adding only a fixed number of intermediate rows (see Figure~\ref{fig:Weakly Constrained Array}). The number, $Z$, of rows to be added (including the target row) solely depends on the parameters of the Markov chain. The arrangement of vertices we would like in the target row is $\sigma_{\text{left}}(U_\pi)$, which is simply $U_\pi$ cyclically shifted to the left by one position. With this as the bottom row of the array $W$, we can stitch together the columns consecutively in the order of their appearance in the array to form a codeword in $S(G)$ of length $N = (m+Z)n+1$, in which each edge $e$ of $G$ appears exactly $P(e)(N-1)$ times. Our main results are the following theorem and its corollary.

\begin{theorem} With $G$ and $P$ as above, let $p = \max\bigl(P(00),P(11)\bigr)$ and $q = \min\bigl(P(00),P(11)\bigr)$. Let $\mathbf{r}$ be any $n$-tuple consisting of $n_1$ occurrences of $v_1$ and $n_2$ occurrences of $v_2$ in some arbitrary order. We can construct a $(G,P,n)$-array with $Z+1$ rows in which the first row is $\mathbf{r}$ and the last row is $\sigma_{\mathrm{left}}(U_\pi)$, where 
$$Z=\max\Bigl(2\bigl(2+\bigl\lceil {\textstyle \frac{P(10)+q-p}{p}} \bigr\rceil\bigl),2\bigl(1+\bigl\lceil {\textstyle \frac{q}{2P(01)}}\bigr\rceil\bigl)\Bigl).$$
\label{thm:Z}
\end{theorem}

We remark that the statement of the theorem will hold for any choice of the last row that has $n_1$ occurrences of $v_1$ and $n_2$ occurrences of $v_2$. 

\begin{corollary}
    Given an $n$-integral stationary Markov chain $P$ on a primitive subgraph, $G$, of $D_{1,2}$, there exists a constant $Z$ that depends only on $P$, such that one can injectively encode $m$ messages from $C_{\mathrm{rbr}}$ into a length $N=(m+Z)n+1$ codeword $w\in S(G)$, in which every pattern $e\in E$ occurs exactly $P(e)(N-1)$ times.  
    \label{cor:Z}
\end{corollary}

Since the corollary is a direct consequence of the discussion prior to the statement of Theorem~\ref{thm:Z}, we will devote the remainder of this section to a proof of the theorem. Leveraging the symmetry of the graph $D_{1,2}$, we may assume that $P(00) \ge P(11)$. Due to space constraints, we provide only a sketch of the proof here, leaving the details to Appendix~\ref{sec:proofdetails}.

\subsection{Proof Sketch for Theorem~\ref{thm:Z}}
Our problem is one of transitioning from a top row $\r = (r^j)_{j\in[n]}$ to a bottom row $\sgl(U_\pi) =: \s = (s^j)_{j\in[n]}$ in a fixed number of steps. If $r^j = v_l$ and $s^j = v_{l'}$ for some $j \in [n]$ and $l,l' \in \{1,2\}$, then the transition from $r^j$ to $s^j$ must be made via a path from $v_l$ to $v_l'$ in $G$. We call such a transition a \emph{$v_l$--$v_{l'}$ flow} (see Figure~\ref{fig:Weakly Constrained Array}), and the number of $v_l$--$v_{l'}$ flows needed to transition from $\r$ to $\s$ is denoted by $M_{v_lv_{l'}}$.
Since $M_{10}+M_{11}$ is the number of flows starting from a vertex $1$ (i.e., $v_2$) in $\r$, we must have $M_{10} + M_{11} = n_2$, which is the number of occurrences of $v_2$ in $\r$. Similarly, considering that flows must terminate in vertices in $\s$, we have $M_{01} + M_{11} = n_2$. So, we have $M_{01} = M_{10} = n_2 - M_{11}$. Also, the equality $M_{00} + M_{01} = n_1$ yields $M_{00} = n_1 - n_2 + M_{11}$. Thus, setting $M_{11} = M$, we can express all the other $M_{**}$ in terms of $M$. This means that we can parametrize our problem by the number $M$.

We divide our problem into two cases: $M\geq P(11)n$ and $M < P(11)n$. For the case $M \geq P(11)n$, our problem can be solved by adding a single intermediate row, when $M$ satisfies a further lower bound (see Eq.~\eqref{eq:LB}). Otherwise, when $M$ does not meet the lower bound in \eqref{eq:LB}, we propose a technique called ``$1$--$1$ boosting'' to raise the number of $1$--$1$ flows to a value $M' > M$. We use this technique repeatedly till we achieve the lower bound. When $M < P(11)n$, we perform $1$--$1$ boosting repeatedly to raise the number of $1$--$1$ flows to a value $M'\ge P(11)n$, so that we are back to the previous case.

Before proceeding further, we introduce some convenient notation. Let $n^* := P(11)n$ denote the multiplicity of the $11$ edge. The multiplicities of the other edges in $G$ can be expressed in terms of $n^*$ as follows: $P(01)n = P(10)n = n_2 - n^*$ and $P(00)n = n_1-(n_2-n^*)$. 

\medskip

\textbf{Case 1: $M \ge n^*$.}\ \ 
\begin{table*}[t]
\begin{center}
  \begin{tabular}{|c|c|c|c|c|c|} 
\hline
  \textbf{Flow} & \textbf{$\#$ occurrences of the flow} & \textbf{Path $\mathbf{1}$} & \textbf{$\#$ times Path $\mathbf{1}$ is used} & \textbf{Path $\mathbf{2}$} & \textbf{$\#$ times Path $\mathbf{2}$ is used} \\
  \hline
   $0$--$0$ &$n_{1}-( n_{2}-M)$ & $0\rightarrow 0 \rightarrow 0 $ &$n_{1} - 2n_{2}+M + n^{*}$ &$ 0\rightarrow 1 \rightarrow 0 $ &$n_{2}-n^{*}$ \\ 
  \hline
  $1$--$0$ &$n_{2}-M$ & $1\rightarrow 0 \rightarrow 0 $ &$n_{2}-M$  & $ 1\rightarrow 1 \rightarrow 0$ &Zero  \\ 
  \hline
  $0$--$1$ &$n_{2}-M$ & $0\rightarrow 0 \rightarrow 1 $ &$n_{2}-M$  & $0\rightarrow 1 \rightarrow 1 $ &Zero\\ 
  \hline
  $1$--$1$ & $M$ & $1\rightarrow 1 \rightarrow 1 $ &$n^{*}$ & $ 1\rightarrow 0 \rightarrow 1 $ &$M - n^{*}$ \\ 
  \hline
\end{tabular}  
\caption{Two-step transition} \label{table:twostep}
\end{center}
\end{table*}
The basic idea here is a two-step transition that attempts to go from $\r$ to $\s$ in two steps, i.e., via one intermediate row. Recall that when the number of $1$--$1$ flows is $M$, the numbers of $0$--$0$, $0$--$1$ and $1$--$0$ flows are $n_1-n_2+M$, $n_2-M$ and $n_2-M$, respectively. (In particular, this means that $M \ge n_2-n_1$ must hold, as the number of $0$--$0$ flows cannot be negative.) 

\subsubsection{Two-step transition} \label{sec:twostep}
The two-step transition method is summarized in Table~\ref{table:twostep}. For each type of flow, Table~\ref{table:twostep} gives two paths of length $2$ that can realize that flow, along with the number of times that each of these paths must be used. Thus, for example, to realize the $n_1-(n_2-M)$ instances of the $0$--$0$ flow that occur in the transition from $\r$ to $\s$, we must use Path~1 ($0 \to 0 \to 0$) exactly $n_1-2n_2+M+n^*$ times, and Path~2 ($0 \to 1 \to 0$) exactly $n_2-n^*$ times. The middle vertices of these paths are the vertices that fill up the intermediate row of the two-step transition. This results in a $3 \times n$ array $W'$ with first row $\r$, last row $\s$, and an intermediate row constructed via Paths~1 and 2 of the flows, \emph{provided that the expressions for the usage numbers for each path are all non-negative and do not exceed the multiplicities of the edges that are in that path}. In particular, we must have 
$n_{1}-2n_{2}+M+n^* \geq 0$ and $n_2 - M \leq P(00)n$. Both conditions reduce to the lower bound
\begin{equation}
  M \geq P(11)n + P(10)n - P(00)n,  
  \label{eq:LB}
\end{equation}
which gives a sufficient condition for the two-step transition method to work. When this condition is met, the path usage numbers given in Table~\ref{table:twostep} ensure that the resulting $3 \times N$ array $W'$ is a valid $(G,P,n)$-array.

\subsubsection{$1$--$1$ boosting}
In problem instances where $M$ does not meet the lower bound in \eqref{eq:LB}, we implement a method of increasing $M$ by augmenting the overall number of steps required for the $\r$-to-$\s$ transition. We call this process of raising the number ($M$) of $1$--$1$ flows as ``$1$--$1$ boosting''. This technique involves creating an intermediate transition problem by introducing a new row, $\r'$, below the top row $\r$ and another new row, $\s'$, above the target bottom row $\s$. We refer to adding one row below $\r$ as the ``launch step'' and one above $\s$ as the ``landing step''. 

Our $1$--$1$ boosting technique is described by the three steps below. An example illustrating this procedure is  provided in Appendix~\ref{sec:11boosting}. 

Step $1$: Start by filling the entries in $\r'$ and $\s'$ corresponding to the $1$--$1$ flows in the $\r$-to-$\s$ transition. Use the $11$ edge $n^*$ times in the launch step and in the landing step; and use the $10$ edge (resp.\ $01$ edge) in the launch step (resp. landing step) $M-n^{*}$ times.

Step $2$: Next, fill the entries in $\r'$ and $\s'$ corresponding to the $0$--$0$ flows. Use the $01$ edge $P(01)n$ times in the launch step, and use the $10$ edge in the landing step for the columns in which the $01$ edge is used in the launch step.

Step $3$: Fill the remaining vacant columns in the launch and landing steps, exhausting all the remaining available edges at both steps.

Let $M'$ be the number of times the $1$--$1$ flow occurs in the intermediate $\r'$-to-$\s'$ transition problem. A small calculation shows that 
    \begin{equation}
   M'=M +P(00)n.
\end{equation}
Thus, if the $1$--$1$ boosting method is iteratively used $t$ times, then at the end of the $t$ iterations, the number of times the \mbox{$1$--$1$} flow occurs in the intermediate transition problem is $M'=M+tP(00)n$. We will use this in Appendix \ref{sec:Steps}, where we compute the overall number of steps it takes to solve the $\r$-to-$\s$ transition problem.

\medskip

\textbf{Case 2: $M < n^*$}. \  We defer this case to Appendices~\ref{sec:case2} and \ref{sec:Steps}.

\begin{example}
   For the graph and the Markov chain from Example~\ref{ex:graphG}, we give a $7 \times 8$ array $W$ that encodes $m=2$ messages using our row-by-row coding scheme.
   
   \begin{center}
\begin{tabular}{ | m{0.5 cm} | m{0.5cm}| m{0.5cm} |m{0.5cm} |m{0.5cm} |m{0.5cm} |m{0.5cm} |m{0.5cm} | } 
  \hline
 0& 0& 0& 0& 0& 0& 1& 1 \\ 
  \hline
  1 & 0 & 0 & 1& 0& 0& 0& 0\\ 
  \hline
  0 & 0 & 0& 0& 0& 0& 1& 1 \\ 
  \hline
   0 & 0 & 1& 0& 1& 0& 0& 0 \\ 
  \hline
   0 & 0 & 0& 0& 0& 1& 1& 0 \\ 
  \hline
   0 & 0 & 0& 1& 1& 0& 0& 0 \\ 
  \hline
   0 & 0 & 0& 0& 0& 1& 1& 0 \\ 
   \hline
\end{tabular}
\end{center}

The first row of the array is $U_{\pi}$, followed by $m=2$ rows corresponding to the payload, and $Z=4$ additional rows added to execute the transition from the third row, $\r$, to the target row $\s = \sgl(U_\pi)$. Note that, the target row $\s$ is actually reached in two steps from $\r$, but the decoder expects a fixed number ($Z=4$) of additional rows as per the coding scheme summarized in Theorem~\ref{thm:Z}. So, we take two more steps to transition from $\s$ to itself. This transition is accomplished by the two-step method described in Section~\ref{sec:twostep}. 

The columns of $W$ can be concatenated in the order of their appearance, i.e.,
\begin{center}
    $w=W_{1}||W_{2}||W_{3}||W_{4}||W_{5}||W_{6}||W_{7}||W_{8}$.
\end{center}
The resulting $w$ is a codeword in $S(G)$ of length $N=(Z+m)n+1 = 49$ that does not have $11$ as a substring, but in which $00$, $01$ and $10$ appear as a substring exactly $24$, $12$ and $12$ times, respectively.
\end{example}

\section{Concluding Remarks}
\label{sec:Conclusion}
In this paper, we proposed a modification of the weakly constrained coding scheme of Buzaglo and Siegel \cite{bornholt2016dna}. We argued that our modified coding scheme is more error-resilient, and also introduces less redundancy, than the original scheme. However, our scheme can only be applied to weak constraints defined by Markov chains on primitive subgraphs of the first-order de Bruijn graph $D_{1,2}$. 
Work is in progress on extending our techniques so as to be applicable to weak constraints defined on higher-order de Bruijn graphs.

\section*{Acknowledgement}
This work was supported in part by a Qualcomm Innovation Fellowship awarded to the second author.


\enlargethispage{-1.2cm} 
\IEEEtriggeratref{6}
\bibliographystyle{ieeetr}
\bibliography{ref}

\newpage

 \newpage

\appendices

\section{Row-By-Row Constrained Coding}
\label{sec:Crbr}
The row–by–row encoding scheme encodes a sequence of $m$ messages $M_{1}, M_{2},\ldots, M_{m}$ into $m$ rows below the $k$ rows of the $U_{\pi}$ matrix defined in section \ref{sec:rowbyrow}. This gives us a resultant $(G,P,n)$-array $W$ of order $(k+m)\times n$, with row index $i$ running from $-k+1 \leq i \leq m $ and column index $j \in [n] $.\\
The messages are encoded into rows such that the following conditions are satisfied:\\
(i) Every message $M_{i} :i \in [m] $ is encoded into a unique $n$ length vector, which is stored in the row $W_{i}$. \\
(ii) For $i_{1},i_{2} \in [m]$, if $i_{1}\leq i_{2} $ then, $W_{i_{1}}$ will be encoded before $W_{i_{2}}$ . \\
(iii) For every edge $e \in E $ and $\forall i \in [m]$, the edge $e$ appears as a column of the rows $ W_{i-k},W_{i-k+1},\ldots, w_{i} $ exactly $ P(e)n$ times.\\

To implement the row-by-row coding, we encode the input messages using a constant-weight code of the form,
\begin{equation}
 C_{\text{rbr}} = C_{1}\times C_{2} \times \ldots \times  C_{|V|} 
\end{equation}

where, 
\begin{equation}
 C_{l} = \{ w \in \Sigma^{n_{l}} :\forall \alpha \in \Sigma , \alpha \text{ appears in $w$ } P(v_{l}\alpha) n \text{ times}\}
\end{equation}

If $M_{i}$ is encoded into a codeword $ \mathbf{c} \in C_{rbr}: \mathbf{c}=c_{1}c_{2} \ldots c_{|V|}$, where $c_{l}\in C_{l} : \forall l \in [|V|]$, then $c_{l}$ is stored in $W_{i}$ ($i^{th}$ row of $W$), in the  $n_{l} $ positions of $j$ for which $W_{i-k,j}, W_{i-k+1,j},\ldots, W_{i-1,j}=v_{l}$. This implementation satisfies the properties of the row-by-row weakly constrained coding scheme.\\

\begin{example} 
\label{ex:Crbr}
Let $G$ be the graph shown in Figure~\ref{fig:No 11constraint system}, representing the "no-$11$" constrained system. $G$ is a primitive subgraph of $D_{1,2}$. Let the $n$-integral stationary Markov chain on $G$ is $P(00)= 0.5, P(01) = P(10) = 0.25$ and $n=8$. Hence, the state probability mass function is $\pi(0)= 0.75, \pi(1)=0.25$, and $n_{0}=6, n_{1}=2$.
\begin{figure}[hbt!]
    \centering
    \includegraphics[width=0.2\textwidth]{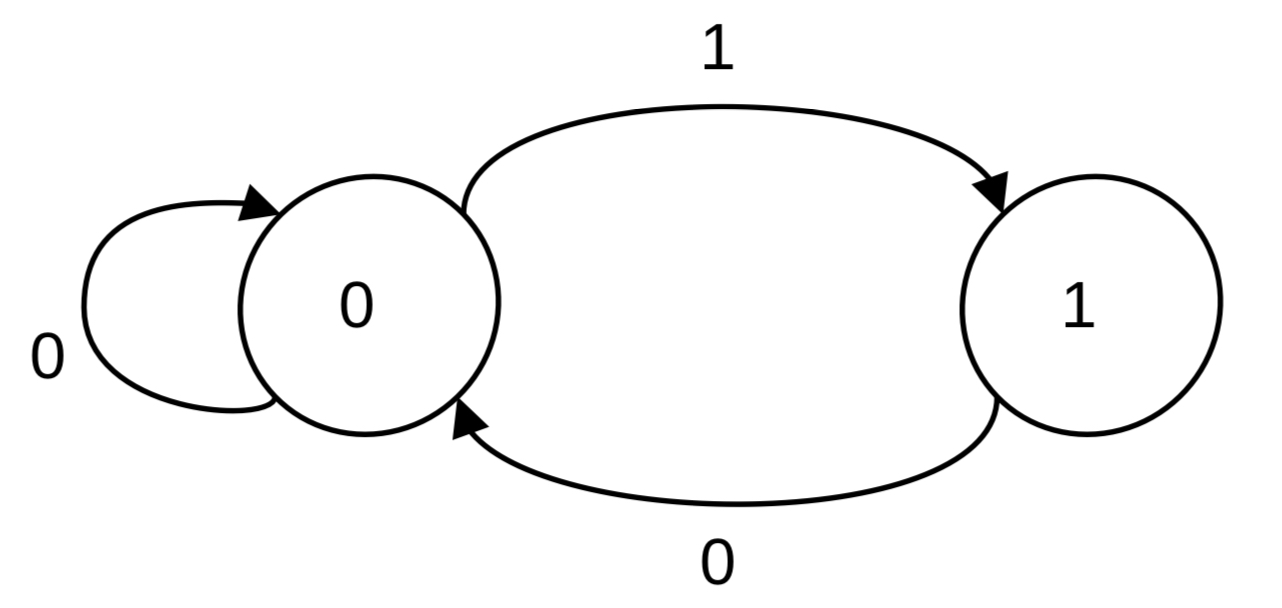}
    \caption{The graph $G=(V,E,L)$}
    \label{fig:No 11constraint system}
\end{figure}
According to the description given by the row-by-row coding scheme, $U_{\pi} = 0 0 0 0 0 0 1 1 $. The constant weight codeword, $C_{\text{rbr}} = C_{1} \times C_{2}$\\

$C_{1} =x \in \{ 0,1 \} ^{6} : \begin{tabular}{l}\\
$1$  appears in $x$ two times \\
$0$ appears in $x$ four times 
\end{tabular}$\\

$C_{2} =  x \in \{ 0,1 \} ^{2} : \hspace{0.2cm}0 $ appears in $x$ two times\\

The codewords $c=10010000,\hat{c}=00001100 \in C_{rbr} $ are encoded into the $W$ matrix. 
\begin{center}
\begin{tabular}{ | m{0.5 cm} | m{0.5cm}| m{0.5cm} |m{0.5cm} |m{0.5cm} |m{0.5cm} |m{0.5cm} |m{0.5cm} | } 
  \hline
 0& 0& 0& 0& 0& 0& 1& 1 \\ 
  \hline
  1 & 0 & 0 & 1& 0& 0& 0& 0\\ 
  \hline
  0 & 0 & 0& 0& 0& 0& 1& 1 \\ 
  \hline
\end{tabular}
\end{center}

The first row is $U_{\pi}$. In the second row, the first six entries of codeword $c$ are encoded below zeroes in the row $W_{0}$, and the last two entries of the codeword $c$ are encoded below ones in $W_{0}$. In the third row, the first six entries of codeword $\hat{c}$ are encoded below zeroes in the row $W_{1}$ and the last two entries of the codeword $\hat{c}$ are encoded below ones in $W_{1}$. Every column belongs to the constrained code, i.e., every column follows the "no-11 constraint". Every edge $e\in E: \{0,1\}^{2}$ occurs as a column of two consecutive rows, exactly $P(e)n$ times.
\label{ex:rbr_encoding}
\end{example}
\section{Incorrect Payload extraction}\label{ex: Error_ex}
\label{sec:Appendix A}
\begin{example} 
The $\hat{U}_\pi$ for the $9\times 8$ array $W$ obtained in Example \ref{ex:graphG} are tabulated below.
\begin{center}
\begin{tabular}{ | m{2 cm} | m{2 cm}| } 
  \hline
 Columns& Markers \\ 
  \hline
  $B_{1}$& $00000$ \\ 
  \hline
 $B_{2}$& $00001$ \\ 
  \hline
 $B_{3}$& $00010$ \\ 
  \hline
 $B_{4}$& $00101$ \\ 
  \hline
$B_{5}$& $01000$ \\ 
  \hline
$B_{6}$& $01010$ \\ 
  \hline
$B_{7}$& $10100$ \\ 
  \hline
 $B_{8}$& $10000$ \\ 
  \hline
\end{tabular}
\end{center}
If $W_{4,5}$ and $W_{4,6}$ entries of the $W$ matrix are flipped, it results in an exchange of the markers for columns $B_{5}$ and $B_{6}$, leading to an erroneous extraction of the payload. Consequently, $c'=10100000, \hat{c}'=00001100 $ (see the correct codeword in Example \ref{ex:rbr_encoding}). If a single entry $W_{2,1}$ is flipped, the column markers for $B_{1}$ and $B_{5}$ become identical, leading to two possible reconstructions of the payload.
\label{ex:Errors}
\end{example}

\section{Proof details}
\label{sec:proofdetails}
\subsection{Example of $1$--$1$ Boosting in the Case of $M \ge n^*$}
\label{sec:11boosting}
\begin{example}Let $G$ be the first order de Bruijn graph shown in figure \ref{fig:Graph}. Consider the $n$-integral stationary Markov chain on $G$  given by $P(10)=P(01)=0.4, P(00)=P(11)=0.1$. For $n=10$, we have $n_{1}=5 \text{ and } n_{2}=5,\text{ and } n^{*}=P(11)n=1$. The initial two-step problem instance is shown in Table \ref{tab:Example 5(a)}. We have two $1$--$1$ flows in our initial two-step problem instance. The transition method given in Table \ref{table:twostep} will be valid only if $M\geq \max( P(11)n,P(11)n + P(10)n - P(00)n)=4$.  As $M \geq n^{*}$, we employ $1$--$1$ boosting for this case. The intermediate transition problem after one iteration of the $1$--$1$ boosting procedure is displayed in Table \ref{tab:Example 5(b)}. After boosting, we obtain three $1$--$1$ flows in the intermediate transition problem. We can repeat this process for one more iteration to raise the number of $1$--$1$ flows to $4$.\\

\begin{table}[h]
\captionsetup{font=scriptsize}
    \centering
    \begin{tabular}{|c|c|c|c|c|c|c|c|c|c|}
         \hline
         $1$ & $1$ & $1$ &$0$ &$1$ & $1$ &$0$ &$0$ &$0$ & $0$ \\
         \hline
           & & & & & & & & &\\
           \hline
         $0$ & $0$ & $0$ &$0$ &$1$ & $1$ &$1$ &$1$ &$1$ & $0$ \\
         \hline
    \end{tabular}
    \small
     \caption{Before $1$--$1$ boosting} 
    \normalsize
    \label{tab:Example 5(a)}
\end{table}

\begin{table}[h]
\captionsetup{font=scriptsize}
    \centering
    \begin{tabular}{|c|c|c|c|c|c|c|c|c|c|}
         \hline
         $1$ & $1$ & $1$ &$0$ &$1$ & $1$ &$0$ &$0$ &$0$ & $0$ \\
         \hline
         $0$ & $0$ & $0$ &$1$ &$1$ & $0$ &$1$ &$0$ &$1$ & $1$ \\
         \hline
           & & & & & & & & &\\
           \hline
           $1$ & $1$ & $0$ &$1$ &$1$ & $0$ &$0$ &$0$ &$0$ & $1$ \\
         \hline
         $0$ & $0$ & $0$ &$0$ &$1$ & $1$ &$1$ &$1$ &$1$ & $0$ \\
         \hline
    \end{tabular}
    \small
     \caption{After one iteration of $1$--$1$ boosting} 
    \normalsize
    \label{tab:Example 5(b)}
\end{table}
\label{ex:11 boosting case1}
\end{example}

\subsection{$1$--$1$ Boosting in the Case of $M < n^*$}
\label{sec:case2}
In the instances where $M < n^{*}$, we perform $1$--$1$ boosting to raise the number of $1$--$1$ flows $M$, to a value $M'\geq n^{*}$. This case is handled in two ways.\\ When $P(11)n-M \geq P(10)n$, perform the following steps : \\
Step 1: Start by filling entries in $\r'$ and $\s'$ corresponding to $1$--$1$ flow in $\r$ to $\s$ transition. Use $11$ edge, $M$ times in the launch step and the landing step.\\
Step 2: Next, fill the entries in $\r'$ and $\s'$ corresponding to $1$--$0$ flow in $\r$ to $\s$ transition with the remaining $11$ edges ($P(11)n-M$) the launch step and use $10$ edge in the landing step.\\
Step 3: Fill the entries in $\r'$ and $\s'$ corresponding to $0$--$1$ flow in $\r$ to $\s$ transition with $01$ edge in the launch step and the remaining $11$ edges ($P(11)n-M $) in the landing step.\\
Let $M'$ be the number of times $1$--$1$ flow occurs in the intermediate $\r'$ to $\s'$ transition problem. A small calculation shows that
\begin{equation}
    M'= M + 2P(10)n. 
\end{equation}

If $1$--$1$ boosting method is iteratively used $t$ times, then at the end of $t$ iterations, the number of $1$--$1$ flow occurs in the intermediate transition problem will be $M'=M+2tP(10)n$. We will use $1$--$1$ boosting for multiple iterations until we achieve $M'\geq n^{*}$.\\

When $P(11)n-M < P(10)n$, perform step $1$ and $2$ in the above defined procedure. The number of $1$--$1$ flows in the intermediate $\r'$ to $\s'$ problem instance will be,
 
\begin{equation}
    M'= M + (P(11)n-M)=P(11)n. 
\end{equation}
Since we have already raised the number of $1$--$1$ flows in the intermediate transition problem to $n^{*}$, we will use this procedure for a single iteration and then use "$11$" boosting for the case $1$.\\

\begin{example}
Let $G$ be the first order de Bruijn graph shown in figure \ref{fig:Graph}. Consider the $n$-integral stationary Markov chain on $G$ given by $P(10)=P(01)=0.1, P(00)=P(11)=0.4$. For $n=10$, we have $n_{1}=5 \text{ and } n_{2}=5$ and $n^{*}=P(11)n=4$. The two-step transition method given in Table \ref{table:twostep} will be valid only if $M \geq \max( P(11)n, P(11)n + P(10)n - P(00)n) = 4$. The initial two-step problem instance is shown in Table \ref{tab:Example 6a}. We have one $1$--$1$ flow in our initial two-step problem instance. As $M < n^{*} \text{ and }P(11)n-M>P(10)n$, we employ $1$--$1$ boosting for the case $2$. The $W$ array after one iteration of the $1$--$1$ boosting is displayed in Table \ref{tab:Example 6b}. After the boosting, we obtain three $1$--$1$ flows. We can repeat this process for one more iteration to raise the number of $1$--$1$ flows to $4$.
\begin{table}[h]
\captionsetup{font=scriptsize}
    \centering
    \begin{tabular}{|c|c|c|c|c|c|c|c|c|c|}
         \hline
         $0$ & $1$ & $1$ &$1$ &$1$ & $0$ &$0$ &$0$ &$0$ & $1$ \\
         \hline
           & & & & & & & & &\\
           \hline
         $0$ & $0$ & $0$ &$0$ &$1$ & $1$ &$1$ &$1$ &$1$ & $0$ \\
         \hline
    \end{tabular}
    \small
    \caption{Before $1$--$1$ boosting} 
    \normalsize
    \label{tab:Example 6a}
\end{table}

\begin{table}[h]
\captionsetup{font=scriptsize}
    \centering
    \begin{tabular}{|c|c|c|c|c|c|c|c|c|c|}
         \hline
         $0$ & $1$ & $1$ &$1$ &$1$ & $0$ &$0$ &$0$ &$0$ & $1$ \\
         \hline
         $0$ & $0$ & $1$ &$1$ &$1$ & $1$ &$0$ &$0$ &$0$ & $1$ \\
         \hline
           & & & & & & & & &\\
           \hline
           $0$ & $0$ & $0$ &$1$ &$1$ & $1$ &$0$ &$1$ &$1$ & $0$ \\
         \hline
         $0$ & $0$ & $0$ &$0$ &$1$ & $1$ &$1$ &$1$ &$1$ & $0$ \\
         \hline
    \end{tabular}
    \small
    \caption{After one iteration of $1$--$1$ boosting}
    \normalsize
    \label{tab:Example 6b}
\end{table}
\label{ex:11 boosting case2}
\end{example} 

\subsection{Number of Steps Required to Reach the Target Row}
\label{sec:Steps}
As our problem is the one of transitioning from a top row $\r = (r^j)_{j\in[n]}$ to a bottom row $\sgl(U_\pi) =: \s = (s^j)_{j\in[n]}$ in a fixed number of steps. In this section, initially, we compute the required number of steps to solve $\r$ to $\s$ transition problem, for each specified case. Subsequently, to ensure a fixed number of steps, $Z$, across all cases, we determine the maximum steps required among all cases. If the transition is achieved before the maximum number of steps, we maintain the target row for the remaining steps. Maintaining the arrangement of vertices after every two steps is possible by choosing $M=n_{2}$ in the two-step transition method described in Table \ref{table:twostep}. Since the maximum of steps needed for each case is an even number, we can maintain the permutation in jumps of two.\\

$\mathbf{\textbf{Case }1:M \geq n^{*}}$ \\

When $M < P(11)n +P(10)n-P(00)n$, let $t$ be the minimum number of iterations of $1$--$1$ boosting needed to raise the number of $1$--$1$ flows in the intermediate $\r'$ to $\s'$ transition problem to a value $M'$ greater than or equal to the further lower bound (see Eq.~\eqref{eq:LB}), then we have \\
\begin{center}
     $M' \geq P(11)n + P(10)n -P(00)n,$ \\
    \vspace{0.3 cm}
$M+tP(00)n \geq P(11)n + P(10)n-P(00)n,$\\
\end{center}

\begin{equation}
t=\left\lceil \frac{P(11)n+P(10)n-P(00)n-M}{P(00)n}\right\rceil.
\end{equation}

Therefore, when the number of $1$--$1$ flows for $\r$ to $\s$ transition is greater than or equal to $n^{*}$ then, for any $n$-integral stationary Markov chain on $G$, it is feasible to have $\r$ to $\s$ transition, in at most $Z_{1}$ steps, where
\begin{equation}
    Z_{1}=2\left(1+\left\lceil \frac{P(11)+P(10)-P(00)}{P(00)}\right\rceil \right).
\end{equation}\\

\textbf{ Case 2 : $\mathbf{M<n^{*}}$}\\

(i) $P(11)n-M \geq P(10)n :$\\

Let $t$ be the minimum number of iterations of $1$--$1$ boosting needed to raise the number of $1$--$1$ flows in the intermediate $\r'$ to $\s'$ transition problem, to a value $M'\geq n^{*}$, then we have
\begin{center}
      $M' \geq P(11)n,$ \\
    \vspace{0.3 cm}
$M+2tP(01)n \geq P(11)n,$\\
\end{center}
\begin{equation}
    t=\left\lceil \frac{P(11)-M}{2P(01)n} \right\rceil. 
\end{equation}
Therefore, when the number of $1$--$1$ flows in $\r$ to $\s$ transition is less than $n^{*}$, then for any $n$-integral stationary Markov chain on $G$ such that $P(11)n-M \geq P(10)n$, it is feasible to have $\r$ to $\s$ transition, in at most $Z_{2}$ steps, where\\

\begin{equation}
    Z_{2}=2\left(1+\left\lceil \frac{P(11)}{2P(01)}\right\rceil \right).
\end{equation}\\

(ii) $P(11)n-M < P(10)n :$\\

We execute the transition from $\r$ to $\s$ in two stages. The first stage is to use one iteration of $1$-$1$ boosting for case when $M<n^{*}$. This increases the number of $1$--$1$ flows in the intermediate $\r'$ to $\s'$ transition to $M'=n^{*}$. The second stage is to use $1$--$1$ boosting for the case when $M\geq n^{*}$ and increase the number of $1$-$1$ flows above the further lower bound, $M'\geq P(11)n+P(10)n-P(00)n$.\\

The first stage can be completed in two steps. Let $t$ be the number of iterations of $1$--$1$ boosting needed to complete the second stage, then we have\\

\begin{center}
    $ M' + tP(00)n \geq P(11)n+P(10)n-P(00)n$,\\
    \vspace{0.3cm}
    $t=\left\lceil \frac{P(11)n+P(10)n-P(00)n-M'}{P(00)n}\right\rceil$.\\ 
\end{center}

Therefore, when the number of $1$--$1$ flows in $\r$ to $\s$ transition is less than $n^{*}$, then for any $n$-integral stationary Markov chain on $G$ such that $P(11)n-M < P(10)n$, it is feasible to have $\r$ to $\s$ transition, in at most $Z_{3}$ steps, where\\

\begin{equation}
    Z_{3}=2\left(2+\left\lceil \frac{P(11)+P(10)-P(00)}{P(00)} \right\rceil\right).
\end{equation}\\

Combining the cases $1\text{ and } 2$, we conclude that for any $n$-integral stationary Markov chain on $G$, it is feasible to have $\r$ to $\s$ transition, in at most $Z$ steps, where $Z=\max(Z_{1}, Z_{2}, Z_{3})$.\\

\newpage
\section{Encoding and Decoding Algorithms}

\begin{algorithm}
\caption{Encoding}
We have a graph $G$ which is a primitive subgraph of $D_{1,2}$, and an $n$-integral stationary Markov chain on $G$ that assigns positive probability to all the edges in $G$.
\begin{algorithmic}[1]
\State Encode the $M$ messages into constant weight codewords, using Row-by-Row coding.
\State Construct $1\times n$ array $U_{\pi}$, as described in section \ref{sec:rowbyrow}.
\State Program $M$ rows below $U_{\pi}$ to obtain $W_{(1+m) \times n}$ array, as per row-by-row coding scheme.
\State Add two vacant rows and set the bottom row as the target row, which is $U_\pi$ cyclically shifted to the left by one position.
\State Compute $M$ the number of $1$--$1$ flows for the two-step transition instance, the lower bound $LB=P(11)n+P(10)n-P(00)n$, and the number of steps $Z$, from the expression given in the Theorem \ref{thm:Z}.
\If {$M \geq \max(P(11)n, LB$)}
\begin{itemize}
    \item [i:] Use the transition method defined in Table \ref{table:twostep}.
    \item [ii:] If the number of steps used is less than $Z$, maintain the target row for the remaining steps by choosing $M=n_{2}$ in the transition method defined in Table \ref{table:twostep}. 
\end{itemize}

    \EndIf
   \If{$P(11)n \leq M \leq LB$}
\begin{itemize}
    \item[i] Use $1$-$1$ boosting for the case $M \geq P(11)n$, and raise the number of $1$-$1$ flows in the intermediate transition instance to a value $M^{'} \geq LB$.
    \item[ii] Apply step $6$.
\end{itemize}
   
     \EndIf
\If{$M < P(11)n \text{ and } P(11)n-M < P(10)n$}
\begin{itemize}
    \item[i] Use the Step $1$ and $2$ of $1$--$1$ boosting for the case $M<P(11)n$ for one iteration.
    \item[ii] Apply step $6$ or $8$.
\end{itemize}
   
     \EndIf     
\If{$M < P(11)m \text{ and } P(11)n-M \geq P(10)n$}
\begin{itemize}
    \item[i]  Use $1$--$1$ boosting for the case $M<P(11)n$ and raise the number of $1$-$1$ flows in the intermediate transition instance to a value $M' \geq n^{*}$.
    \item[ii] Apply step $6$ or $8$.
\end{itemize}
     \EndIf
   \State Concatenate the columns of $W$ array in the order of their appearance to get one long codeword of length $N$ in which every pattern $e\in E$ occurs exactly $P(e)(N-1)$ times.
\end{algorithmic}
\end{algorithm}
\begin{algorithm}[!h]
\caption{Decoding}
\begin{algorithmic}[1]
\State Reconstruct $W$ array by stacking each batch of $Z$ symbols as a column in the order of their appearance.
\State Use the row-by-row decoding method to extract the payload.
\end{algorithmic}
\end{algorithm}

\end{document}